\documentclass[prl,twocolumn,aps,superscriptaddress,showpacs,floatfix]{revtex4}
\usepackage{amsmath}
\usepackage{graphicx}
\newcommand{\R}{{\cal R}}
\renewcommand{\v}{{\rm vac}}
\begin{document}
\title{Surface Josephson plasma waves in layered superconductors}
\author{Sergey Savel'ev}
\affiliation{Frontier Research System, The Institute of Physical
and Chemical Research (RIKEN), Wako-shi, Saitama, 351-0198, Japan}

\author{Valery Yampol'skii}
\affiliation{Frontier Research System, The Institute of Physical
and Chemical Research (RIKEN), Wako-shi, Saitama, 351-0198, Japan}

\affiliation{A. Ya. Usikov Institute for Radiophysics and
Electronics, NASU, 61085 Kharkov, Ukraine}

\author{Franco Nori}
\affiliation{Frontier Research System, The Institute of Physical
and Chemical Research (RIKEN), Wako-shi, Saitama, 351-0198, Japan}

\affiliation{Physics Department, MCTP, CSCS, University of
Michigan, Ann Arbor, MI 48109-1040, USA}
\date{\today}

\begin{abstract}
We predict the existence of surface waves in layered
superconductors in the THz frequency range, below the Josephson
plasma frequency $\omega_J$. This wave propagates along the
vacuum-superconductor interface and dampens in both transverse
directions out of the surface (i.e., towards the superconductor
and towards the vacuum). This is the first prediction of
propagating surface waves in any superconductor. These predicted
surface Josephson plasma waves are important for different
phenomena, including the complete suppression of the specular
reflection from a sample (Wood's anomalies) and a huge enhancement
of the wave absorption (which can be used as a THz detector).
\end{abstract}
\pacs{74.72.Hs
} \maketitle

High temperature $\rm Bi_2Sr_2CaCu_2O_{8+\delta}$ superconductors
have a layered structure of superconducting CuO$_2$ planes with
Josephson coupling between them~\cite{reviews1}. This structure
favors the propagation of electromagnetic waves, called Josephson
plasma waves (JPW)~\cite{plasma-theor,plasma-exp,tachiki}, through
the layers. These waves attract considerable interest because of
their Terahertz frequency range (from 300 GHz to 30 THz
corresponding to 1000-90 $\mu$m (wavelength)), which is still
hardly reachable for both electronic and optical
devices~\cite{tera-appl}. During the last decade there have been
many attempts to push THz science and technology forward
\cite{tera-appl} because of many important applications in
physics, astronomy, chemistry, biology and medicine, including THz
imaging, spectroscopy, tomography, medical diagnosis, health
monitoring, environmental control, as well as chemical and
biological identification.

The unusual optical properties of layered superconductors,
including reflectivity and transmissivity, caused by the JPW
excitation were studied in, e.g.,
Refs.~\cite{plasma-theor,opt1,helm1,helm,crist}. In particular,
Ref.~\cite{helm1} demonstrated that the spectrum of JPW consists
of two branches due to a peculiar effect~\cite{break} of dynamical
breaking of charge neutrality and, therefore, the transmissivity
exhibits a sharp peak at frequencies just above the Josephson
plasma frequency $\omega_J$.

All previous works on this problem have focused on running waves
in the frequency range {\it above} the gap of the JPW spectrum,
i.e., above the Josephson plasma frequency $\omega>\omega_J$. A
similar gapped spectrum also appears in solid state
plasma~\cite{platzman}. In such situations the presence of the
sample boundary can produce a new branch of the wave spectrum
inside the gap, i.e., a surface plasmon~\cite{platzman,agr}. In
general, surface waves play a very important role in many
fundamental resonant phenomena, such as the ``Wood anomalies'' in
the reflectivity~\cite{agr,wood} and transmissivity~\cite{wood1}
of periodically corrugated metal and semiconducting samples, and
are employed in many devices. Therefore, it is important to know
if surface waves can exist in layered superconductors.

Here, we prove that {\it surface} Josephson plasma waves can
propagate along the surface between the superconductor and the
vacuum in a wide frequency range below $\omega_J$. We derive the
dispersion relation for these waves and propose ways to excite
these.
We show that these surface waves play an important role in the
absorption and reflection of electromagnetic waves, including
their resonance dependence on the incident angle $\theta$. The
studied resonant absorbability could be experimentally observed by
measurement of the surface impedance of a sample or dc
resistivity. The predicted phenomena are potentially useful for
designing THz detectors.

\begin{figure}[!htp]
\begin{center}
\hspace*{-0.6cm}\includegraphics*[width=10cm]{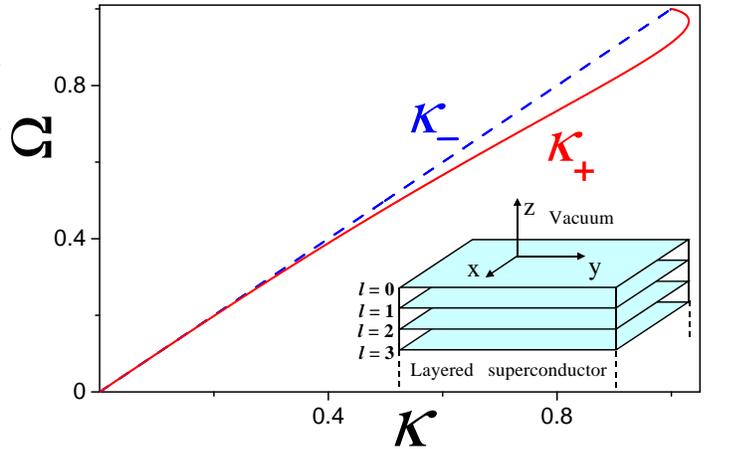}
\vspace*{5.5cm}
 \caption{(Color online) Spectra
of the surface Josephson plasma waves $\kappa_{\mp}$ 
for the parameters $\alpha=0.1,$ and $\beta=1.4$, standard for
Bi-2212. The inset shows the geometry used.}\label{f1}
\end{center}
\end{figure}

{\it Surface Josephson plasma waves.---} We consider a
semi-infinite layered superconductor in the geometry shown in the
inset in Fig.~\ref{f1}. Using a standard approximation, the
spatial variations inside the very thin superconducting layers are
neglected in the direction perpendicular to the layers. The
crystallographic $\mathbf{ab}$-plane coincides with the
$\mathbf{xy}$-plane and the $\mathbf{c}$-axis is along the
$z$-axis. Superconducting layers are numbered by the integer
$l\geq 0$.

We consider a surface $p$--wave with the electric, ${\vec E} =
\{E_x, 0, E_z\}$, and magnetic, ${\vec H} = \{0, H, 0\}$, fields
damped away from the interface $z=0$,
\begin{equation}
H,\, E_x,\, E_z  \propto \exp(-{ i}\omega t + {  i}qx - klD)
\label{super}
\end{equation}
inside a layered superconductor, $z<0$, and
\begin{equation}\label{vac}
H^\v,\, E_x^\v,\, E_z^\v \propto \exp(-{  i}\omega t + { i}qx -
k_\v z),
\end{equation}
in vacuum, for $z>0$ and $q>\omega/c$. Here $D$ is the spatial
period of the layered structure.

The Maxwell equations for waves (\ref{vac}), $qH^\v = - (\omega/c)
E_z^\v, \ k_\v H^\v = -({  i}\omega/c) E_x^\v, \ \ { i}qE_z^\v +
k_\v E_x^\v = -({  i}\omega/c) H^\v,$
 provide both the usual dispersion relation,
$ k_\v=\sqrt{q^2-\omega^2/c^2},$ for the wave in vacuum and the
ratio of amplitudes for tangential electric and magnetic fields at
the interface $z=+0$ (i.e., right above the sample surface)
\begin{equation}\label{e41}
\frac{E_x^\v}{H^\v}= \frac{{  i}c}{\omega}
\sqrt{q^2-\omega^2/c^2}.
\end{equation}

Inside the layered superconductor, where $z<0$, the gauge
invariant phase difference is described by a set of coupled
sine-Gordon equations. For Josephson plasma waves, these
non-linear equations can be linearized and rewritten in terms of
the magnetic fields $H^l$ between the $l$th and $(l+1)$th layers,
\[
\left(1-\frac{\lambda_{ab}^2}{D^2}\partial^2_l\right)\left(\frac{\partial^2
H^{l}}{\partial t^2}+ \omega_r \frac{\partial H^{l}}{\partial t}
+\omega_J^2 H^{l}\right.
\]
\begin{equation}\label{e62}
\left. -\alpha \omega_J^2\partial^2_lH^{l}\right)-
\frac{c^2}{\varepsilon}\frac{\partial^2 H^{l}}{\partial x^2}=0.
\end{equation}
Here $\lambda_{ab}$ is the London penetration depth in the
$z$-direction, the operator $\partial_l^2$ is defined as
$\partial_l^2 f_l = f_{l+1}+f_{l-1}-2f_l$, $\omega_J = \sqrt{8\pi
e D J_c/\hbar\varepsilon}$ is the Josephson plasma frequency
determined by the maximum Josephson current $J_c$ and dielectric
constant $\varepsilon$, and $\omega_r$ is the relaxation
frequency. The effect of breaking charge neutrality \cite{break},
which is crucial for our analysis, is taken into account in
Eq.~(\ref{e62}). The constant $\alpha$ characterizing this effect
was estimated, e.g., in Ref.~\cite{helm}, $\alpha \sim
0.05$--$0.1$ for Bi-2212 or Tl-2212 crystals.

Substituting the wave (\ref{super}) in Eq.~(\ref{e62}), we obtain
the implicit equation for the damped-wave transverse wave vector
$k(q,\omega)$
\[
\frac{\omega^2}{\omega^2_J}=1
+\frac{\lambda_c^2q^2}{1-(4\lambda_{ab}^2/D^2)\sinh^2[k(q,\omega)D/2]}
\]
\begin{equation}\label{e46}
-4\alpha\sinh^2[k(q,\omega)D/2], \qquad
\lambda^2_c=\frac{c^2}{\omega_J^2\varepsilon}
\end{equation}
for $\omega_r\rightarrow 0$. A similar spectrum, but for running
Josephson plasma waves (with imaginary $k(q,\omega)$), was earlier
obtained in Ref.~\cite{helm}. At $\omega<\omega_J$, solving
Eq.~(\ref{e46}) with respect to $\sinh^2[k(q,\omega)D/2]$ results
in two branches of positive transverse spatial decrement,
$k_\pm(q,\omega)>0$. It is important to stress that
Eq.~(\ref{e46}) is not a spectrum of the studied {\it
longitudinal} surface waves. Indeed, Eq.~(\ref{e46}) has two free
parameters, $\omega$ and $q$. We need to obtain the $q(\omega)$
dispersion relation for our surface waves.
This dispersion relation $q(\omega)$ can be obtained by joining
fields (\ref{super}) in a superconductor and (\ref{vac}) in vacuum
at the sample surface via the boundary conditions. Thus, in order
to find the spectrum of the surface Josephson plasma waves, we can
derive the ratio $E_x/H$ in the superconductor at the sample
surface and equate this ratio to Eq.~(\ref{e41}).

The difference between the magnetic field $H^\v$ in vacuum and the
value $H^{0}$ between the 0th and the 1st superconducting layers
is described by the London equation,
\begin{equation}\label{e43}
\frac{H^\v-H^{0}}{D}\ \approx\ \frac{A_{x0}}{\lambda_{ab}^2} \
\approx \ \frac{-{  i}c}{\lambda_{ab}^2 \omega}E_{x0}.
\end{equation}
Here $A_{x0}$ and $E_{x0}$ are the $x$-components of the vector
potential and electric field in the first superconducting layer.
We ignore the displacement current in Eq.~(\ref{e43}) because it
is proportional to a small parameter $(\lambda_{ab} \omega /c)^2$.
Besides, we neglected the contribution of the gauge-invariant
scalar potential into the $x$-component of the electric field
since the Debye length is much shorter than the wave length
$q^{-1}$. In order to eliminate $H^0$, we use the relation $
H^\v-H^{0}\;=\;H^\v[1-\exp(-kD)],$  that follows from
Eq.~(\ref{super}) with $l=0$ and $l=1$. Now, using
Eq.~(\ref{e43}), we obtain the ratio between electric and magnetic
fields at $z=-0$ (i.e., right below the sample surface)
\begin{equation}\label{e45}
\frac{E_x}{H^\v}= \frac{{  i}\omega \lambda_{ab}^2}{cD}
[1-\exp(-kD)].
\end{equation}

Using the continuity conditions for the tangential components of
electric field at the surface and Eqs.~(\ref{e41}), (\ref{e46}),
and (\ref{e45}), we obtain the dispersion relation for two
branches of the surface wave corresponding to two solutions of
Eq.~(\ref{e46}). For $(1-\omega/\omega_J)\; \gg\;
\sqrt{\alpha/\varepsilon}D/\lambda_{ab}\; \approx\; 5\cdot
10^{-4}$, this spectrum can be written as
\begin{eqnarray}\label{e52}
\kappa_-(\Omega) = \Omega; \ \ \ \ \ \ \kappa_+(\Omega) =
\Omega\biggl(1+\beta^2\Omega^2\Gamma_{\Omega}
\nonumber \\
\times 
\left\{1+2\Gamma_{\Omega}\left[ 1 -
\left(1+\Gamma_{\Omega}^{-1}\right)^{1/2} \right]
\right\}\biggr)^{1/2}
\end{eqnarray}
for two branches ``$\mp$". Here we introduce the dimensionless
variables: $\kappa =cq/\omega_J, \qquad \Omega = \omega/\omega_J,
\qquad \beta = 2\lambda_{ab}^2 \omega_J/cD$, and
$\Gamma_{\Omega}=(1-\Omega^2)/4\alpha$. The value of the parameter
$\beta$ for Bi-2212 is about 1.4. The spectra Eqs.~(\ref{e52}) are
shown in Fig.~\ref{f1}. Both branches merge in a narrow frequency
region below $\omega_J$, i.e.,
$(1-\Omega)\;\sim\;(\alpha/\varepsilon)^{1/2}(D/\lambda_{ab})$.

The surface mode $\kappa_-(\Omega)$ attenuates into the vacuum at
very large distances, of about $c\lambda_{ab}/\omega D$, and seems
to be difficult to observe experimentally. Another mode,
$\kappa_+(\Omega)$, dampens on scales $\sim\ c/\omega$ and is of
significant interest. As we show below, it plays an important role
in the transmissivity and reflectivity properties of layered
superconductors at frequencies $\omega<\omega_J$.

{\it Excitation of the surface waves; resonant electromagnetic
absorption.---} One of the ways to excite surface waves is via
externally applied electromagnetic waves on a sample having
spatially modulated parameters. Thus, we consider a weak
modulation of the maximum current density $J_c$ via, say, creation
of pancake vortices by the out-of-plane magnetic field. This can
result in the modulation of $\omega_J$. For simplicity, we assume
that
\begin{equation}
\omega_J(x)^2=\omega_J^2\left[1+2\mu\cos\left(\frac{2\pi
x}{a}\right)\right], \ \ \mu\ll 1,
\end{equation}
where $a$ is a spatial period.

An electromagnetic wave with $\omega<\omega_J$ incident at an
angle $\theta$ with respect to the sample surface generates modes
having longitudinal wave vectors $q_m=\omega\sin\theta/c+2\pi
m/a$, with integer $m$. Almost all of these modes for $m\neq 0$
are weak, because $\mu\ll 1$. However, one of these modes (e.g.,
for $m=1$) can be excited with large amplitude at resonance, i.e.,
when the wave vector $q_1=(\omega/c)\sin\theta+2\pi/a$ is close to
the wave vector $q_+=\omega_J\kappa_+(\Omega)/c$ of the surface
wave (\ref{e52}). This corresponds to the incident angle $\theta$
close to the resonance angle $\theta_0$ defined by
\begin{equation}
\Omega\sin\theta_0+\frac{2\pi c}{a\omega_J}=\kappa_+(\Omega).
\end{equation}
Because of this resonance, the amplitude of the wave with $q=q_1$
can be of the same order, or even higher than, the amplitude
$H_{\rm in}^\v$ of the incident wave. In resonance, the mode with
$q=q_1$ is actually the surface Josephson plasma wave discussed
above.

Let us discuss the mechanism of excitation of surface waves. The
incident $H_{\rm in}^\v$ (red arrow in Fig.~2) and
specularly-reflected $H_{\rm sp}^\v$ (blue dashed arrow in Fig.~2)
electromagnetic waves
\begin{equation}
H_{\rm in;sp}^{\rm vac}\exp\left\{ {i}\left(\frac{x\sin\theta\pm
z\cos\theta}{c}-t \right)\omega\right\}
\end{equation}
generate the wave (in green, Fig.~2) damped inside the
superconductor,
\begin{equation}
H_{0}\exp\left\{ {i}\left(\frac{x\sin\theta}{c}-t
\right)\omega-\kappa_-\!\!\left(\frac{\omega\sin\theta}{c},\omega\right)Dl\right\},
\label{h0}
\end{equation}
having a lower $\kappa_-(\Omega)$, compared to $\kappa_+(\Omega)$.
Indeed, as was shown in Ref.~\cite{helm}, the longer wave is
mainly generated at $(\omega-\omega_J)/\omega_J\; \gg\;
\sqrt{\alpha/\varepsilon}D/\lambda_{ab}\; \approx\; 5\cdot
10^{-4}$. Moreover, at low frequencies, this longer wave exhibits
a correct limiting behavior, producing magnetic fields described
by the well-known London equation. The waves $H_{\rm in}^{\rm
vac}$, $H_{\rm sp}^{\rm vac}\approx H_{\rm in}^{\rm vac}$, and
$H_0\approx 2H_{\rm in}^{\rm vac}$ represent the solution of the
problem in the zero approximation with respect to $\mu$.

To first order approximation, the solution of Eq.~(\ref{e62}) with
$\omega_J(x)$ inside the superconductor consists of both the
forced component,
\begin{equation}
H_{\rm forced}\exp\left\{ {i}q_1x- {i}\omega
t-\kappa_-\!\!\left(\frac{\omega\sin\theta}{c},\omega\right)Dl\right\},
\end{equation}
which attenuates with the same decrement as $H_0$, and free
oscillations
\begin{equation}
H_{\rm surf}\exp\left\{ {i}q_1x- {i}\omega
t-\kappa_+\left(q_1,\omega\right)Dl\right\}.
\end{equation}
The amplitude $H_{\rm forced}$ is defined by Eq.~(\ref{e62})
itself, whereas the amplitudes, $H_{\rm surf}$ (in dotted magenta)
and corresponding vacuum mode
\begin{equation}
H_{\rm surf}^{\rm vac}\exp\left\{ {i}q_1x- {i}\omega
t-\left(q_1^2-\frac{\omega^2}{c^2}\right)^{1/2}z\right\},
\end{equation}
are determined by the boundary conditions for harmonics $q=q_1$ of
$H$ and $E_x$ at $z=0$. Solving the corresponding set of
nonhomogeneous linear equations for $H_{\rm surf}$ and $H_{\rm
surf}^{\rm vac}$ we obtain the amplitude of the resonance wave:
\begin{equation}
H_{\rm surf}^{\rm vac}=H_{\rm in}^{\rm vac}\frac{2\mu
q_0^2}{(1-\Omega^2)(q_1^2-q_0^2)}\cdot\frac{1}{\R}
\end{equation}
with the denominator
\begin{equation}
\R=X+
{i}Y=\frac{2\kappa_+(\Omega)\cos\theta_0(\theta-\theta_0)}{\beta\Omega\sqrt{\kappa_+^2(\Omega)
-\Omega^2}}+\frac{
{i}\omega_r\alpha\Omega^3}{\omega_J(1-\Omega^2)^3}.
\end{equation}

\begin{figure}[!htp]
\begin{center}
\includegraphics*[width=8.0cm]{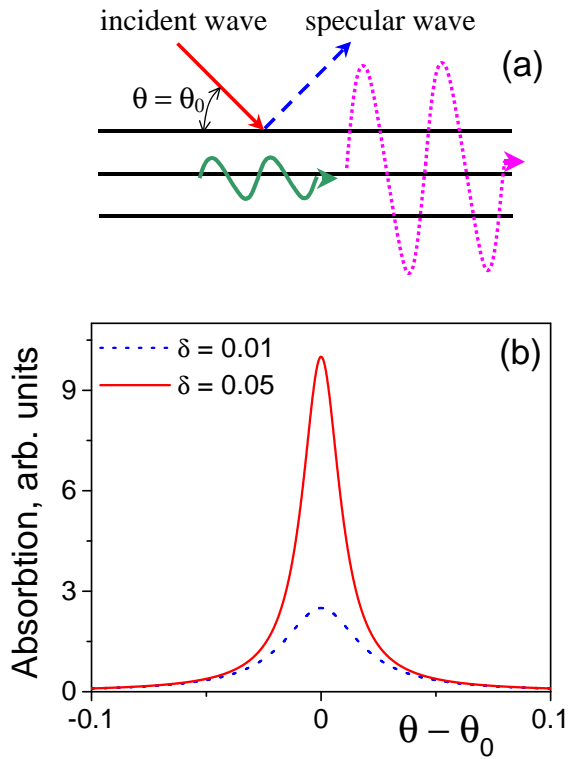}
\vspace{-0.5cm}
 \caption{(Color online) (a) Schematic diagram showing the mechanism
of excitation of the surface waves along the superconductor-vacuum
interface. To zero-order approximation, with respect to the
amplitude $\mu$ of the spatial modulations in a superconducting
sample, an incident wave (shown as a solid red arrow) reflects as
a specular wave (the straight dashed blue arrow), producing a
damped wave (green wave) inside superconductors. To first order
approximation, the very intense surface  wave (dotted magenta
wave) can be excited at a certain resonant angle between the
incident wave and sample surface. (b) Absorption obtained using
Eq.~(\ref{absorb}) for different effective dampings
$\delta=\beta\omega_r\alpha\Omega^4(k_+^2-\Omega^2)^{1/2}\,[2\omega_J(1-\Omega^2)^3k_+\cos\theta_0]^{-1}$.}
 \label{f2}
\vspace{-0.5cm}
\end{center}
\end{figure}

The resonance is characterized by the Lorentz form
$1/\R=1/[X(\theta)+ {i}Y]$. The excitation of the surface wave
results in the resonant peak in the electromagnetic absorption,
\begin{equation}
{\rm Absorption}(\theta)\ \propto \ \sigma_{\perp}
E_z^2+\sigma_{\parallel} E_x^2\ \propto\
\frac{1}{X^2(\theta)+Y^2}, \label{absorb}
\end{equation}
shown in Fig.~\ref{f2}. The resonance in the absorption can be
observed by measuring the dependence of the surface impedance on
the angle $\theta$. Alternatively, the peak in absorption produces
a temperature increase, resulting in a sharp increase of the DC
resistance or even the transition of the sample to the normal
state at $\theta=\theta_0$.

Naturally, the absorption peak is accompanied by the resonant
decrease of the amplitude $H_{\rm sp}^{\rm vac}$ of the specularly
reflected wave. Even though this effect is of second-order with
respect to $\mu$, it can result in a significant (up to complete)
suppression of $H_{\rm sp}^{\rm vac}$ at $\theta=\theta_0$ due to
the resonance denominator $\R(\theta)$. In the optical range, this
phenomenon is known as Wood's anomalies \cite{wood} of
reflectivity and has been used for several diffraction devices.

{\it THz detectors.---} These effects could be potentially useful
for the {\it design of THz detectors}, an important current goal
of many labs worldwide. The simplest design could be a spatially
modulated Bi2212 sample fixed on a precisely rotated holder and
attached by contacts to measure its resistance. Spatial
modulations in the sample could be fabricated by either using ion
irradiation of the sample covered by periodically modulated mask
\cite{kwok}, or even mechanically \cite{grooves}. When rotating
the sample, the incident THz radiation can produce a surface wave
at certain angles. This results in a strong enhancement of
absorption associated with increasing of temperature in the sample
and, thus, its resistance. The relative positions of the resonance
peaks (the set of angles) allows to calculate {\it the angle and
the frequency of the incident THz radiation}, while the relative
heights of the resistance peaks can be used to estimate {\it the
intensity of the incident radiation.}\/

{\it Conclusions.---} We have derived the surface Josephson plasma
waves in layered superconductors and obtained their dispersion
relation. The absorption of the incident electromagnetic wave can
strongly increase at certain incident angles due to the resonant
generation of the predicted surface waves. This is the first
prediction of propagating surface waves in any superconductor. We
propose a way to experimentally observe these surface waves.

We acknowledge partial support from the NSA and ARDA under AFOSR
contract No. F49620-02-1-0334, and by the NSF grant No.
EIA-0130383.

\end{document}